\documentclass[aps,prl,twocolumn,superscriptaddress,showpacs]{revtex4}
\usepackage{amsmath,amssymb,graphicx}
\usepackage[hypertex,colorlinks=true,citecolor=blue,urlcolor=blue]{hyperref}

\newcommand{\PR}[4]{
Phys. Rev. #1 {\bf #2}, \href{http://dx.doi.org/10.1103/PhysRev#1.#2.#3}{#3} (#4)}
\newcommand{\PRRC}[4]{
Phys. Rev. #1 {\bf #2}, \href{http://dx.doi.org/10.1103/PhysRev#1.#2.#3}{#3(R)} (#4)}
\newcommand{\PRL}[3]{
Phys. Rev. Lett. {\bf #1}, \href{http://dx.doi.org/10.1103/PhysRevLett.#1.#2}{#2} (#3)}
\newcommand{\RMP}[3]{
Rev. Mod. Phys. {\bf #1}, \href{http://dx.doi.org/10.1103/RevModPhys.#1.#2}{#2} (#3)}
\newcommand{\JPSJ}[3]{
J. Phys. Soc. Jpn. {\bf #1}, \href{http://dx.doi.org/10.1143/JPSJ.#1.#2}{#2} (#3)}
\newcommand{\Nature}[4]{
Nature (London) {\bf #2}, \href{http://dx.doi.org/10.1038/#1}{#3} (#4)}
\newcommand{\Nat}[5]{
Nat. #2 {\bf #3}, \href{http://dx.doi.org/10.1038/#1}{#4} (#5)}

\begin{document}
\title{
Laser-Driven Multiferroics and Ultrafast Spin Current Generation}

\author{Masahiro Sato}
\affiliation{Department of Physics, Ibaraki University, 
Mito, Ibaraki 310-8512, Japan}
\affiliation{Department of Physics and Mathematics, Aoyama-Gakuin University, 
Sagamihara, Kanagawa 229-8558, Japan}
\affiliation{Advanced Science Research Center, Japan Atomic Energy Agency, 
Tokai, Ibaraki 319-1195, Japan}
\affiliation{ERATO, Japan Science and Technology Agency, Sendai, Miyagi 
980-8577, Japan}
\author{Shintaro Takayoshi}
\affiliation{Department of Quantum Matter Physics, University of Geneva,
24 quai Ernest-Ansermet, Geneva 1211, Switzerland}
\author{Takashi Oka}
\affiliation{Max-Planck-Institut f\"{u}r Physik komplexer Systeme
             (MPI-PKS), N\"{o}thnitzer Stra\ss e 38,
             Dresden 01187, Germany}
\affiliation{Max-Planck-Institut f\"{u}r Chemische Physik fester Stoffe
             (MPI-CPfS), N\"{o}thnitzer Stra\ss e 40, Dresden 01187, Germany}

\date{\today}

\begin{abstract}

We propose an ultrafast way to generate spin chirality 
and spin current in a class of multiferroic magnets
using a terahertz circularly polarized laser. 
Using the Floquet formalism for periodically driven systems,
we show that it is possible to dynamically control the Dzyaloshinskii-Moriya 
interaction in materials with magnetoelectric coupling. 
This is supported by numerical calculations, by which additional 
resonant phenomena are found. 
Specifically, when a static magnetic field is applied in addition to the 
circularly polarized laser, a large resonant enhancement of 
spin chirality is observed resembling the electron spin resonance. 
Spin current is generated when the laser is spatially modulated 
by chiral plasmonic structures 
and could be detected using optospintronic devices. 

\end{abstract}


\maketitle

{\it Introduction.}---
Control of emergent collective phenomena by external fields is an 
important problem in condensed matter. 
Multiferroic magnets (for a review, see Refs.~\cite{Wang09,Tokura10,Tokura14})
are opening new possibilities in this direction since the 
local spins are coupled not only to magnetic fields
but to electric fields through the magnetoelectric (ME) coupling. 
Laser control of materials is attracting interest
with a goal of realizing ultrafast and noncontact 
manipulation~\cite{Kirilyuk10,Vicario13,Schellekens14,Choi14,Oka09,
Kitagawa11,Lindner11,Tsuji08,Tsuji09,Jotzu14,Wang13}. 
In the research community of magnetic systems, control of magnetism 
using a laser is being studied in the context of spin-pumping and 
spintronics~\cite{Kirilyuk10,Vicario13,Schellekens14,Choi14}.
On the other hand, in the field of electronic systems, 
periodically driven quantum systems draw the interest of many researchers. 
When the Hamiltonian is time periodic, the system can be described by the 
so-called Floquet states~\cite{Shirley1965,Sambe1973}, 
a temporal analog of the Bloch states, and it is possible 
to control their quantum nature. 
For noninteracting systems, the control of the band topology 
has been studied theoretically~\cite{Oka09,Kitagawa11,Lindner11}
and experimentally~\cite{Jotzu14,Wang13}. 
It is possible to understand the effect of a laser through a mapping from 
the time-dependent Hamiltonian to a {\it static} 
effective Hamiltonian using the Floquet theory, and 
the change of quantum state, e.g., topology and symmetry, 
is attributed to the emergent terms in the static effective Hamiltonian. 
This framework can also be applied to quantum magnets. 
Laser-induced magnetization growth in general 
quantum magnets~\cite{Takayoshi14-1,Takayoshi14-2} 
as well as laser-driven topological spin states~\cite{Takayoshi14-2,Sato14}, 
a quantum spin versions of 
Floquet topological insulators, were proposed recently. 

In the current work, we apply the Floquet theory to 
quantum {\it multiferroics} and study the synthetic interactions appearing
in the effective Floquet Hamiltonian [see Eq.~\eqref{eq:synthetic}].
We show that when elliptically or circularly polarized lasers are 
applied, an additional Dzyaloshinskii-Moriya (DM) 
interaction~\cite{Dzyaloshinsky58} emerges
and its direction (DM vector) can be controlled. 
The DM interaction generally favors a spiral magnetic order
and if its strength is spatially modulated, it is possible to 
induce spin currents. 
Through direct numerical calculations, we verify this picture, 
and then propose a way to generate ultrafast spin currents 
in a realistic device by optical means.

{\it Multiferroics with laser application.}---
In multiferroics~\cite{Tokura14,Tokura10,Wang09}, 
spin degrees of freedom couple to electromagnetic waves
not only through the Zeeman coupling, but also
through the ME coupling. 
This is because the local polarization vector is related to spin degrees 
of freedom from crystallographic reasons. 
The Hamiltonian for multiferroics subject to a laser can be expressed as
\begin{equation}
 {\cal H}(t)={\cal H}_{0}+{\cal H}_{\rm E}(t)+{\cal H}_{\rm B}(t),
 \label{eq:TimeDepHamil}
\end{equation}
where ${\cal H}_{0}$ is the spin Hamiltonian, 
and the laser-driven time-dependent terms ${\cal H}_{\rm E}(t)
=-\boldsymbol{E}(t)\cdot\boldsymbol{P}$ and 
${\cal H}_{\rm B}(t)=-g\mu_{\rm B}\boldsymbol{B}(t)
\cdot\boldsymbol{S}$ respectively denote the ME coupling 
of the total polarization $\boldsymbol{P}$ with electric field 
$\boldsymbol{E}(t)$, and the Zeeman coupling between the total spin
$\boldsymbol{S}$ with the magnetic field $\boldsymbol{B}(t)$ 
($g$ is Land\'e's $g$ factor and $\mu_{\rm B}$ is Bohr magneton). 
The polarization $\boldsymbol{P}$ is given by a function of spin operators. 
Electric and magnetic components of the laser are represented as 
$\boldsymbol{E}(t)=E_{0}(\cos(\Omega t+\delta),
-\sin(\Omega t),0)$ and 
$\boldsymbol{B}(t)=E_{0}c^{-1}(-\sin(\Omega t),
-\cos(\Omega t+\delta),0)$, respectively. The value of $\delta$ fixes 
the helicity of the laser, 
i.e., $\delta=0$, $\pi$, and $\pi/2$ respectively 
corresponds to right-circularly, left-circularly, and 
linearly polarized lasers. Symbols $\Omega$ and $c$ stand for the laser 
frequency and the speed of light, respectively.

{\it Synthetic interactions from Floquet theory.}---
We apply the Floquet theory and the $\Omega^{-1}$ expansion to 
Eq.~(\ref{eq:TimeDepHamil}). 
From the discrete Fourier transform of
the time-periodic Hamiltonian,
${\cal H}(t)=\sum_{m}e^{-im\Omega t}H_{m}$  
($m$: integer), the static effective Hamiltonian 
${\cal H}_{\rm eff}=\sum_{i\ge0}\Omega^{-i}{\cal H}_{\rm eff}^{(i)}$ 
can be expanded in terms of $\Omega^{-1}$ and the leading two terms 
are given by~\cite{Casas2003,Mananga2011,Supple}
\begin{equation}
 {\cal H}_{\rm eff}^{(0)}=H_{0}, \quad
 {\cal H}_{\rm eff}^{(1)}=-\sum_{m>0}[H_{+m},H_{-m}]/m.
\label{eq:FloquetHeff}
\end{equation}
For large enough $\Omega$, 
we can truncate ${\cal H}_{\rm eff}$ up to the $\Omega^{-1}$ order. 
In the present multiferroic system, the first correction 
${\cal H}_{\rm eff}^{(1)}$, which we call the synthetic interaction, 
is given by
\begin{align}
 {\cal H}_{\rm syn}&\equiv
   \Omega^{-1} {\cal H}_{\rm eff}^{(1)} 
   =-\frac{i\cos\delta}{2\Omega}\big\{
     \alpha^{2}[\tilde P^{x},\tilde P^{y}]\nonumber\\
   &+\alpha\beta([\tilde P^{x},S^{x}]+[\tilde P^{y},S^{y}])
     +\beta^{2}[S^{x},S^{y}]\big\}
\label{eq:synthetic}
\end{align}
with $\alpha=g_{\rm me}E_{0}$, and 
$\beta=g\mu_{\rm B}E_{0}c^{-1}$. Here, $g_{\rm me}$ is the ME 
coupling constant [see Eq.~(\ref{eq:KNB})] 
with $\tilde{\boldsymbol{P}}$ being 
a dimensionless function of spins, 
i.e., $\boldsymbol{P}=g_{\rm me} \tilde{\boldsymbol{P}}$.
Let us comment on the magnitude of the synthetic terms. 
The strongest magnetic field $\beta$ of a terahertz (THz) laser 
attains 1--10 T~\cite{Hirori11,Pashkin13}. 
The magnitude of $g_{\rm me}(\Omega)$ can 
be large in a gigahertz (GHz) to THz 
region~\cite{Takahashi12,Cheong09}, 
and from both experimental and theoretical 
analyses~\cite{Takahashi12,Cheong09,Katsura07,Furukawa10}, 
the value of $\alpha$ is expected to be of the same order as $\beta$. 
If we use as reference the typical value of exchange coupling 
$J=0.1$--10 meV ($\sim 1$--100 T) in standard magnets 
(e.g., XXZ magnets in Eq.~\eqref{eq:XXZ}) 
both $\alpha/J$ and $\beta/J$ can achieve values of 0.1--1.

The precise form of the synthetic interaction 
depends on the type of the ME coupling. 
Here we consider the case where the polarization 
$\boldsymbol{P}=\sum_{\boldsymbol{r}, \boldsymbol{r}'} 
\boldsymbol{P}_{\boldsymbol{r}, \boldsymbol{r}'}$
is given by a product of two spin operators on sites 
($\boldsymbol{r}$, $\boldsymbol{r}'$).
$\boldsymbol{P}_{\boldsymbol{r}, \boldsymbol{r}'}$ is proportional to 
the exchange interaction (energy density) 
$\boldsymbol{S}_{\boldsymbol{r}}\cdot\boldsymbol{S}_{\boldsymbol{r}'}$
in symmetric magnetostriction type multiferroics~\cite{Moriya67},
while it is proportional to the vector spin chirality 
$\boldsymbol{S}_{\boldsymbol{r}}\times\boldsymbol{S}_{\boldsymbol{r}'}$
in the antisymmetric magnetostriction type (also known as the inverse 
DM effect)~\cite{Tanabe65,Katsura05,Katsura07,
Mostovoy06,Dagotto06}.
The term $[\tilde P^x,\tilde P^y]$ thereby yields three spin terms
such as the scalar spin chirality. 
In Ref.~\onlinecite{Sato14}, 
it was shown that a three-spin term related to the scalar spin 
chirality is generated in the symmetric ME coupling case and 
can induce a topological gap in spin liquids. 
In addition, $[\tilde{P}^{a},S^{b}]$ and $[S^{x},S^{y}]$ 
induce two-spin and single-spin terms, respectively.

\begin{figure}[t]
\includegraphics[width=0.4\textwidth]{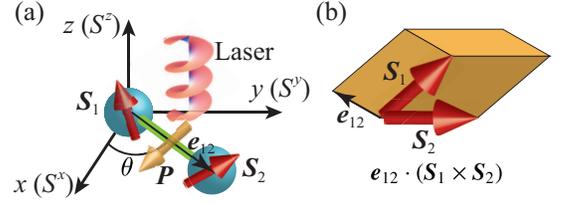}
\caption{(a) Schematic picture of a multiferroic system 
consisting of two spins $\boldsymbol{S}_{1,2}$ 
in a circularly polarized laser. The vector 
$\boldsymbol{P}$ represents the electric polarization. 
(b) Schematic picture for geometric meaning of 
the synthetic DM interaction in Eq.~(\ref{eq:EffHamil}). 
}
\label{fig:Setup}
\end{figure}

{\it Two-spin system.}---
To illustrate the effect of Eq.~(\ref{eq:synthetic}), 
let us first focus on a simple two-spin multiferroic model 
depicted in Fig.~\ref{fig:Setup}(a). 
The applied laser travels toward the $-z$ direction, 
and the two-spin multiferroic magnet is within the $xy$ plane. 
We assume that the two-spin system $\boldsymbol{S}_{1,2}$ possesses 
an electric polarization $\boldsymbol{P}$ through the ME coupling as 
\begin{equation}
 \boldsymbol{P}=g_{\rm me}\boldsymbol{e}_{12}\times
   (\boldsymbol{S}_{1}\times\boldsymbol{S}_{2}),
\label{eq:KNB}
\end{equation}
where $\boldsymbol{e}_{12}=(\cos\theta,\sin\theta,0)$ 
is the vector connecting two spins 
(the distance between spins is set to unity). 
This ME coupling is known to be responsible for electric polarization 
in a wide class of spiral ordered (i.e., chirality ordered) 
multiferroic 
magnets~\cite{Tokura14,Tokura10,Wang09,Katsura05,Mostovoy06,Dagotto06}. 
Using  Eq.~(\ref{eq:synthetic}), we obtain 
the synthetic interaction
\begin{equation}
{\cal H}_{\rm syn}=\frac{\alpha\beta}{2\Omega}\cos\delta
   (\boldsymbol{e}_{12}\cdot\boldsymbol{\cal V}_{12})
   +\frac{\beta^{2}}{2\Omega}\cos\delta(S_{1}^{z}+S_{2}^{z}),
   \label{eq:EffHamil}
\end{equation}
where $\boldsymbol{\cal V}_{12}=\boldsymbol{S}_{1}
\times\boldsymbol{S}_{2}$ is the vector spin chirality. 
The first term is the laser-driven DM interaction and generated 
via the single-photon absorption and emission as shown in 
Ref.~\onlinecite{Supple}. This DM term is geometrically illustrated 
by the volume of a parallelepiped as in Fig.~\ref{fig:Setup}(b). 
The three spin term from $[\tilde P^x,\tilde P^y]$ 
disappears in Eq.~(\ref{eq:EffHamil}) 
since $\boldsymbol{e}_{12}$ is within the polarization plane.

The result~(\ref{eq:EffHamil}) is valid for any spin Hamiltonian 
${\cal H}_{0}$ with arbitrary spin magnitude $S$. 
In the original model~\eqref{eq:TimeDepHamil}, 
the DM vector in $\boldsymbol{E}(t)\cdot\boldsymbol{P}$ is parallel to 
$z$ axis. On the other hand, Eq.~\eqref{eq:EffHamil} shows that 
the synthetic DM vector is in the direction of $\boldsymbol{e}_{12}$, 
which is in the $xy$ plane and perpendicular to the $z$ axis. 
The coefficient $\alpha\beta$ in Eq.~\eqref{eq:EffHamil} indicates that 
both ME and Zeeman terms are necessary 
for emergence of the synthetic DM interaction. 
It is also significant that the laser should be circularly 
or elliptically polarized. 
In fact, ${\cal H}_{\rm syn}$ vanishes
when the laser is linearly polarized ($\delta=\pi/2$). 
We emphasize that the synthetic DM coupling constant and its sign 
can be controlled by changing the laser helicity.

We comment on the importance of breaking the SU(2) symmetry 
of the system. If the spin Hamiltonian ${\cal H}_{0}$ is 
spin-rotationally [i.e., SU(2)] symmetric, 
the Zeeman term ${\cal H}_{\rm B}(t)$ commutes with ${\cal H}_{0}$. 
This means that the laser-driven $\beta$ term 
plays no role in the growth of spin chirality. 
Thus, it is important that the system has 
magnetic anisotropy or spontaneous symmetry breakdown that
relaxes the SU(2) symmetry.

{\it Many-spin system (spirals and chiral-solitons).}--- 
It is straightforward to extend our result~(\ref{eq:EffHamil}) 
to multiferroic magnets consisting of many spins. 
For instance, the static effective 
Hamiltonian for an 1D multiferroic spin chain ${\cal H}_{0}^{\rm 1D}$ 
along the $x$ axis ($\theta=0$) with a circularly polarized laser 
is given by 
\begin{equation}
 {\cal H}_{\rm eff}^{\rm 1D}
   ={\cal H}_{0}^{\rm 1D}
     \pm\sum_{j}\frac{\alpha\beta}{2\Omega}{\cal V}_{j,j+1}^{x}
     \pm\sum_{j}\frac{\beta^{2}}{2\Omega}S_{j}^{z}, 
\label{eq:Eff_chain}
\end{equation}
where 
$\boldsymbol{\cal V}_{j,j+1}\equiv\boldsymbol{S}_{j}\times\boldsymbol{S}_{j+1}$, 
and the sign $\pm$ respectively corresponds to $\delta=0$ and $\pi$. 
Here we assume that the bond polarization $\boldsymbol{P}_{j,j+1}$ is 
proportional to the bond chirality $\boldsymbol{\cal V}_{j,j+1}$, 
and the total polarization is given by 
$\boldsymbol{P}_{\rm tot}=g_{\rm me}\sum_{j}\boldsymbol{e}_{j,j+1}\times
\boldsymbol{\cal V}_{j,j+1}$ ($\boldsymbol{e}_{j,j+1}$ stands for 
a vector connecting the spin site $j$ and $j+1$). 

\begin{figure}[t]
\includegraphics[width=0.4\textwidth]{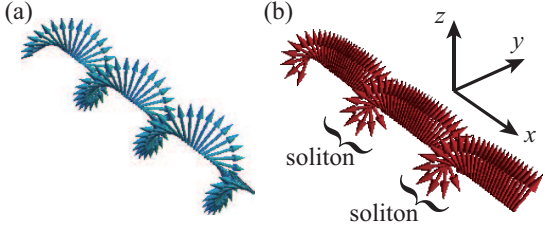}
\caption{(a) Spin spiral (helical) ordered state 
and (b) chiral-soliton-lattice state can be laser-induced 
if the exchange coupling is AF and FM, respectively.
}
\label{fig:Textures}
\end{figure}

The effective model (\ref{eq:Eff_chain})
is known to support interesting spin states with spatial modulations
if the interaction in ${\cal H}_{0}^{\rm 1D}$
is short ranged. 
A spin spiral state [Fig.~\ref{fig:Textures}(a)] emerges 
when the exchange is antiferromagnetic (AF)
due to the competition between exchange and laser-induced DM interaction. 
On the other hand, in the case of a ferromagnetic exchange, it is known 
that competition among exchange, DM and Zeeman couplings can lead to 
a chiral-soliton-lattice state 
[Fig.~\ref{fig:Textures}(b)]~\cite{Kishine05,Togawa13} 
as the classical ground state of ${\cal H}_{\rm eff}^{\rm 1D}$. 
This indicates that a laser can create several types of spiral 
spin textures depending on lattices and interactions of 
the multiferroic system ${\cal H}_{0}$.

{\it Numerical analysis.}---
The Floquet effective Hamiltonian and 
the predicted emergence of the synthetic interaction (\ref{eq:synthetic}) 
are the general result and apply to a broad class of multiferroics. 
However, there are limitations to the theory: (i)
The effective Hamiltonian is applicable when the 
driving frequency (=photon energy) $\Omega$ is much larger 
than all the other energy scales in the system, 
and (ii) when many-body interactions are present, 
the system eventually heats up~\cite{Alessio14,Lazarides14}. 
As a complementary test, we use a numerical approach and
perform direct time dependent calculations 
in a laser-driven multiferroic model based on 
${\cal H}(t)$ (\ref{eq:TimeDepHamil}). Here, we focus on 
simple multiferroic XXZ spin-$\frac{1}{2}$ chains aligned in the $x$ direction 
($\theta=0$) with an external magnetic field $H$
\begin{equation}
{\cal H}_{0}^{\rm 1D}= {\cal H}_{\rm XXZ}=\sum_{j}
(J\boldsymbol{S}_{j}\cdot\boldsymbol{S}_{j+1}
   -J\Delta S_{j}^{x}S_{j+1}^{x}-HS_{j}^{x}).
 \label{eq:XXZ}
\end{equation}
In order to break the SU(2) symmetry, 
we introduced either an Ising anisotropy 
$-J\Delta S_{j}^{x}S_{j+1}^{x}$ or a static Zeeman term $-HS_{j}^{x}$. 
In the case of circularly polarized laser with $\delta=0$ ($\delta=\pi$), 
the effective Hamiltonian Eq.~(\ref{eq:Eff_chain})
predicts the emergence of $x$ component of vector chirality 
$\langle{\cal V}_{\rm tot}^{x}\rangle<0$ $(>0)$.

\begin{figure}[t]
\includegraphics[width=0.48\textwidth]{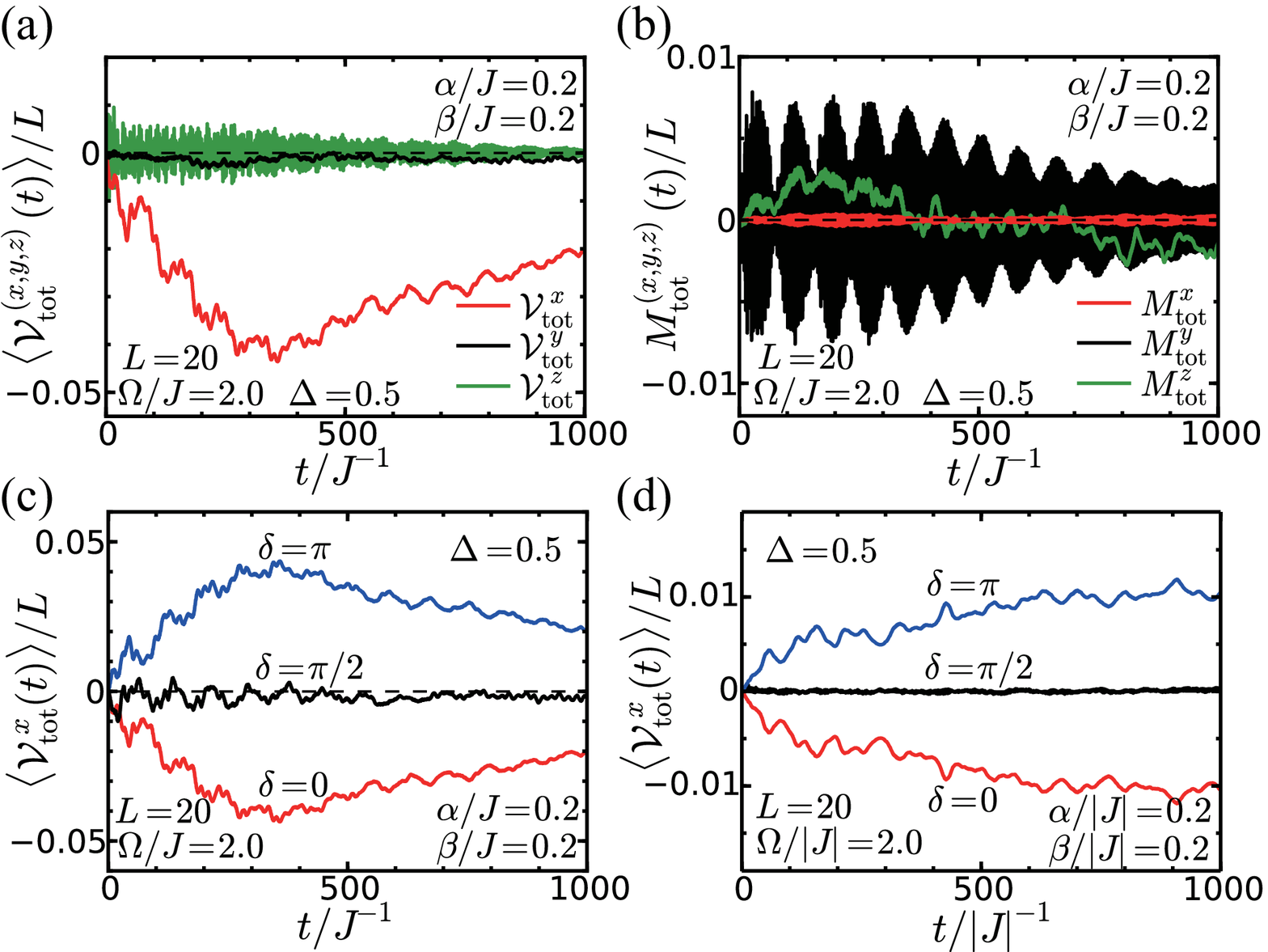}
\caption{Simulation results of 
a multiferroic XXZ chain  ($\Delta=0.5,\;H/J=0$) in 
a circularly polarized laser. 
Time evolutions of (a) vector chirality 
$\langle\boldsymbol{\cal V}_{\rm tot}(t)\rangle$ and 
(b) magnetization $\boldsymbol{M}_{\rm tot}(t)$
in an AF XXZ model under a circularly polarized laser 
($J>0$ and $\delta=0$). 
(c) Laser helicity ($\delta$) dependence of 
$\langle{\cal V}_{\rm tot}^{x}(t)\rangle$. 
(d) Time evolution of vector chirality in the case of a 
ferromagnetic exchange ($J<0$).}
\label{fig:XXZTimeEvol}
\end{figure}

We perform simulations for finite-size systems 
with $L$ spins. The initial state is set to the ground state of 
Eq.~(\ref{eq:XXZ}) obtained by numerical diagonalization. 
The laser is turned on at $t=0$ and 
the system evolves according to the time-dependent Hamiltonian 
${\cal H}(t)$ (\ref{eq:TimeDepHamil}). 
The time evolution of the state $|\Psi(t)\rangle$ 
is obtained by integrating the Schr\"odinger equation 
$i(d/dt)|\Psi(t)\rangle={\cal H}(t)|\Psi(t)\rangle$
using the fifth order Runge-Kutta method. 
In the numerical analysis below, we set $\alpha/J=\beta/J=0.2$.

First, consider the case of $\Delta=0.5$ and $H=0$. 
In Figs.~\ref{fig:XXZTimeEvol}(a) and \ref{fig:XXZTimeEvol}(b), 
we plot the typical time evolutions of vector chirality 
$\langle\boldsymbol{\cal V}_{\rm tot}(t)\rangle=
\langle\sum_{j}\boldsymbol{\cal V}_{j,j+1}(t)\rangle$ and 
magnetization $\boldsymbol{M}_{\rm tot}(t)
=\langle\sum_{j}{\boldsymbol S}_{j}(t)\rangle$ 
for an XXZ model with $H=0$ in a 
circularly polarized laser with $\Omega/J=2$ and $\delta=0$. 
The vector chirality $\langle{\cal V}_{\rm tot}^{x}(t)\rangle<0$ appears 
as expected while $\langle{\cal V}_{\rm tot}^{(y,z)}(t)\rangle$ remains small. 
The dependence of the vector chirality on the laser helicity $\delta$ 
is depicted in Figs.~\ref{fig:XXZTimeEvol}(c) and \ref{fig:XXZTimeEvol}(d). 
We see that $\langle{\cal V}_{\rm tot}^{x}(t)\rangle$ becomes negative (positive)
for $\delta=0$ $(\pi)$, while it remains very small 
for linear polarization $\delta=\pi/2$. 
These behaviors are consistent with the prediction (\ref{eq:Eff_chain}) 
from the Floquet theory. 
However, the vector chirality does not keep on growing 
but becomes saturated around $t/J^{-1}\sim 400$ 
in Fig.~\ref{fig:XXZTimeEvol}(c). This may be due to heating; 
the system's ``effective temperature'' exceeds the magnitude of the
synthetic term ($\sim \alpha\beta/\Omega$) 
around this time, and the linear growth of the chirality stops. 
This is consistent with recent studies on ``heating'' in 
closed periodically driven systems 
that have revealed that the effective Hamiltonian approach, e.g., 
Eq.~(\ref{eq:Eff_chain}), 
is valid only for finite time, and if the driving is continued the system 
will approach an infinite temperature state \cite{Alessio14,Lazarides14}. 
When the system is coupled to a heat reservoir, the heating can 
be stopped and the system can be stabilized \cite{Tsuji09}. 
Figure \ref{fig:XXZTimeEvol}(b) shows that the 
magnetization $\boldsymbol{M}_{\rm tot}(t)$ does not 
grow but only exhibits an oscillation 
with small amplitude.

\begin{figure}[t]
\includegraphics[width=0.48\textwidth]{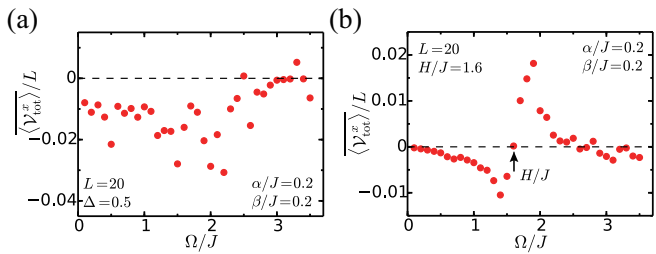}
\caption{$\Omega$ dependence of time average of vector chirality 
$\overline{\langle{\cal V}_{\rm tot}^{x}\rangle}$ 
in (a) AF ($J>0$) 
XXZ model without external magnetic field 
and (b) Heisenberg model ($\Delta=0$) with external magnetic field.
Around $\Omega=H$, we observe a large magnitude of laser-driven 
chirality due to a resonant behavior. 
}
\label{fig:Field}
\end{figure}

In order to understand how the induced chirality 
depends on the laser frequency $\Omega$, 
we define its time average as
\begin{equation}
\overline{\langle\boldsymbol{\cal V}_{\rm tot}\rangle}\equiv
   \frac{1}{T}\int_{0}^{T}dt
   \langle\boldsymbol{\cal V}_{\rm tot}(t)\rangle
 \label{eq:V_TimeAverage}
\end{equation}
with $T=1000 J^{-1}$. As shown in Fig.~\ref{fig:Field}(a),
for the XXZ chain in zero field, the induced chirality is typically negative, 
which agrees with the prediction 
from the Floquet effective Hamiltonian (\ref{eq:Eff_chain}), 
but since Eq.~(\ref{eq:Eff_chain}) is based on the 
high frequency expansion, 
it fails to explain the detailed structure in the simulation. 
We find many small peaks in Fig.~\ref{fig:Field}(a) 
that are presumably due to resonance with many-body excited states.

We also consider laser-driven spin chains in a static magnetic field. 
Naively, we may expect that the magnetic field will 
play the same role as the Ising anisotropy, i.e., 
a source to break the SU(2) symmetry, and 
no qualitative difference would occur. 
However, in Fig.~\ref{fig:Field}(b), 
the result of direct calculation shows 
a resonant behavior in the generation of a vector chirality 
around $\Omega\sim H$, 
which is clearly not described by the effective 
Hamiltonian~(\ref{eq:Eff_chain}). 
We verified that a similar resonant behavior 
also occurs in a multiferroic spin-1 chain and a spin-$\frac{1}2$ ladder 
(see Supplemental Material~\cite{Supple}), which indicates that 
the resonance around $\Omega\sim H$ is universal 
in a broad class of multiferroic systems. 
What happens around $\Omega\sim H$ is analogous to 
electron spin resonance (ESR). Thus, our calculation implies that by 
using circularly polarized laser in an ESR setup, 
it is possible to efficiently generate a vector chirality in multiferroics.


\begin{figure}[t]
\includegraphics[width=0.35\textwidth]{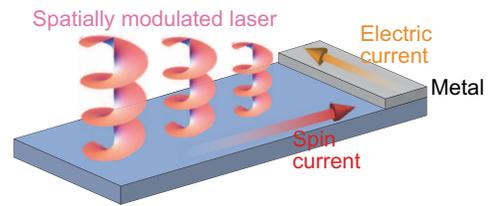}
\caption{Setup to detect signatures of 
laser-driven DM interactions by measuring spin current. 
The spin current pumped from a spin chain to a metal with a strong 
spin-orbit coupling is changed into an electric current via 
inverse spin Hall effect in the metal.}
\label{fig:DetectionDM}
\end{figure}

{\it Detection schemes.}---
Finally, we propose schemes to detect the 
synthetic DM interaction and vector chirality. 
Detection using pump-probe optical methods is in principle possible 
by observing nontrivial textures, e.g., spiral states~\cite{Supple}. 
Another scheme is to utilize optospintronic methods with plasmon resonances 
proposed in Ref.~\cite{Uchida15}. 
The underling idea is to apply a {\it spatially modulated} circularly polarized laser 
to a multiferroic magnet (Fig.~\ref{fig:DetectionDM}). Assuming that the 
exchange coupling is dominant in the time-dependent Hamiltonian of 
the spin chain ${\cal H}_{\rm XXZ}+{\cal H}_{\rm E}(t)+{\cal H}_{\rm B}(t)$, 
the Heisenberg equation of motion shows that 
\begin{equation}
 i(dS_{j}^{x}/dt)
   \approx[S_{j}^{x},{\cal H}_{\rm XXZ}]
   =iJ({\cal V}_{j-1,j}^{x}-{\cal V}_{j,j+1}^{x}).
\end{equation}
Similar expressions hold for higher dimensions and therefore 
 a finite spin current  $\langle dS_{j}^{x}/dt\rangle$ appears due to a site-dependent vector chirality if the laser is spatially modulated. 
A circularly polarized laser with large spatial modulation can be 
realized in the near field of chiral plasmonic structures \cite{Schaferling16}. 
Then, in order to detect the spin current, one needs to 
combine the plasmonic structure with a metallic electrode 
where the spin current is transformed into an electric current 
by inverse spin Hall effect~\cite{Saitoh06,Valenzuela06,Maekawa12,Uchida16}. 
The spin current is injected from the multiferroic magnet to the electrode 
if $\langle dS_{j}^{x}/dt\rangle$ is nonzero 
at the interface~\cite{Tserkovnyak02,Adachi}.
Using materials with strong spin-orbit coupling such as Pt~\cite{Uchida16} for the electrode, 
we can observe the generation of a laser-driven chirality 
through an electric voltage drop. 
In the Supplemental Material~\cite{Supple}, we numerically show that 
an inhomogeneous chirality appears when we apply a spatially modulated laser.

Distinction of mechanisms is an important issue. 
In contrast with other effects of laser such as heating, 
the laser-driven DM interaction strongly depends on the direction of the laser 
as can be seen from Fig.~\ref{fig:Setup} and Eq.~(\ref{eq:KNB}), 
and thus systematic measurements with sample rotation are useful for 
making clear the origin.

{\it Summary.}--- 
In conclusion, we proposed a way to generate and control DM 
interactions and spin currents in multiferroics utilizing 
elliptically polarized lasers. 
Our understanding is based on the Floquet theory 
with the $\Omega^{-1}$ expansion, 
which captures the general tendency of the numerical results, 
while we find an additional resonant enhancement of spin chirality 
when a static magnetic field is applied.

\acknowledgements
We would like to thank Hidekazu Misawa, Yuichi Ohnuma, Stephan Kaiser, Thomas Weiss and 
Shin Miyahara for fruitful discussions. 
M.S. is supported by KAKENHI 
(Grants No. 26870559, No. 25287088, and No. 15H02117), 
S.T. by the Swiss NSF under Division II, 
and T.O. by KAKENHI (Granst No. 23740260 and No. 15H02117). 
S.T. and T.O. are also supproted by ImPact project (No. 2015-PM12-05-01) 
from JST.

\clearpage
\appendix
\renewcommand{\theequation}{S\arabic{equation}}
\setcounter{equation}{0}
\renewcommand{\thefigure}{S\arabic{figure}}
\setcounter{figure}{0}

\begin{widetext}
\begin{center}
 \textbf{\Large Supplemental Material}
\end{center}

\subsection{S1. Floquet theorem and $\Omega^{-1}$ expansion}
Here, we explain the Floquet theorem and the $\Omega^{-1}$ expansion scheme of 
the effective Hamiltonian in relation 
with Eqs.~\eqref{eq:FloquetHeff}, \eqref{eq:synthetic}, and \eqref{eq:EffHamil}
in the main text. This theorem can be viewed as time version of 
Bloch theorem~\cite{Kittel} for quantum systems with 
spatially periodicity (such as crystals).

We consider a time-dependent Schr\"odinger equation 
\begin{equation}
i \frac{\partial}{\partial t}|\Psi(t)\rangle 
= {\cal H}(t)|\Psi(t)\rangle. 
\label{eq:Sch}
\end{equation}
for a time-periodic system ${\cal H}(t)={\cal H}(t+T)$.
We perform discrete Fourier transform for 
the Hamiltonian ${\cal H}(t)=\sum_{m}e^{-im\Omega t}H_{m}$ where 
the frequency $\Omega$ is $2\pi/T$, and $m$ is an integer
running from $-\infty$ to $\infty$. 
The solution can be written as $|\Psi(t)\rangle={\rm e}^{-i\epsilon t}|\Phi(t)\rangle$
where $|\Phi(t)\rangle$ is the Floquet state, 
which is periodic in time, i.e., $|\Phi(t)\rangle=|\Phi(t+T)\rangle$, 
and $\epsilon$ is the Floquet quasienergy. 
The Floquet state can be expanded as
$|\Phi(t)\rangle=\sum_{m}{\rm e}^{-{\rm i}m\Omega t}|\Phi^{m}\rangle$. 
Substituting this to Eq.~(\ref{eq:Sch}), we obtain 
the following eigenvalue equations 
\begin{equation}
\sum_{m}(H_{n-m}-m\Omega\delta_{mn})|\Phi^{m}\rangle
                            =\epsilon|\Phi^{n}\rangle.
\label{eq:FloquetEigen}
\end{equation}
In the model studied in the main text, nonzero components are only 
time-independent part and terms proportional to $\exp(\pm i\Omega t)$, 
hence we can rewrite Eq.~(\ref{eq:FloquetEigen}) 
in the following matrix form:
\begin{equation}
\begin{pmatrix}
\;\ddots&&&&&&\\
&H_{0}-2\Omega&H_{+1}&0&0&0&\\
&H_{-1}&H_{0}-\Omega&H_{+1}&0&0&\\
&0&H_{-1}&H_{0}&H_{+1}&0&\\
&0&0&H_{-1}&H_{0}+\Omega&H_{+1}&\\
&0&0&0&H_{-1}&H_{0}+2\Omega&\\
&&&&&&\ddots\;
\end{pmatrix}
\begin{pmatrix}
\vdots\\
|\Phi^{ 2}\rangle\\
|\Phi^{ 1}\rangle\\
|\Phi^{ 0}\rangle\\
|\Phi^{-1}\rangle\\
|\Phi^{-2}\rangle\\
\vdots
\end{pmatrix}
=\epsilon
\begin{pmatrix}
\vdots\\
|\Phi^{ 2}\rangle\\
|\Phi^{ 1}\rangle\\
|\Phi^{ 0}\rangle\\
|\Phi^{-1}\rangle\\
|\Phi^{-2}\rangle\\
\vdots
\end{pmatrix}.
\label{eq:FloquetEigenMatrix}
\end{equation}
The block structure appearing in Eqs.~(\ref{eq:FloquetEigen}) and 
(\ref{eq:FloquetEigenMatrix}) is understood intuitively as ``photon dressed 
states''. The frequency $\Omega$ is interpreted as a photon energy, 
and thereby $|\Phi^{-m}\rangle$ is regarded as a state in Hilbert subspace 
with $m$ photons. This can also be seen from the energy shift $m\Omega$ 
in the diagonal components of Eq.~(\ref{eq:FloquetEigenMatrix}). 
Different subspaces are hybridized by $H_{+1}$ and $H_{-1}$, which 
correspond to photon emission and absorption processes, respectively. 
In general, $H_{\pm m}$ represents direct multi-photon processes. 
Figure~\ref{fig:Floquet} illustrates the hybridization structure 
of the Floquet system, 
where a step in the Floquet direction corresponds 
to increase or decrease of photon energy.

\begin{figure}[b]
\centering
\includegraphics[width=0.7\textwidth]{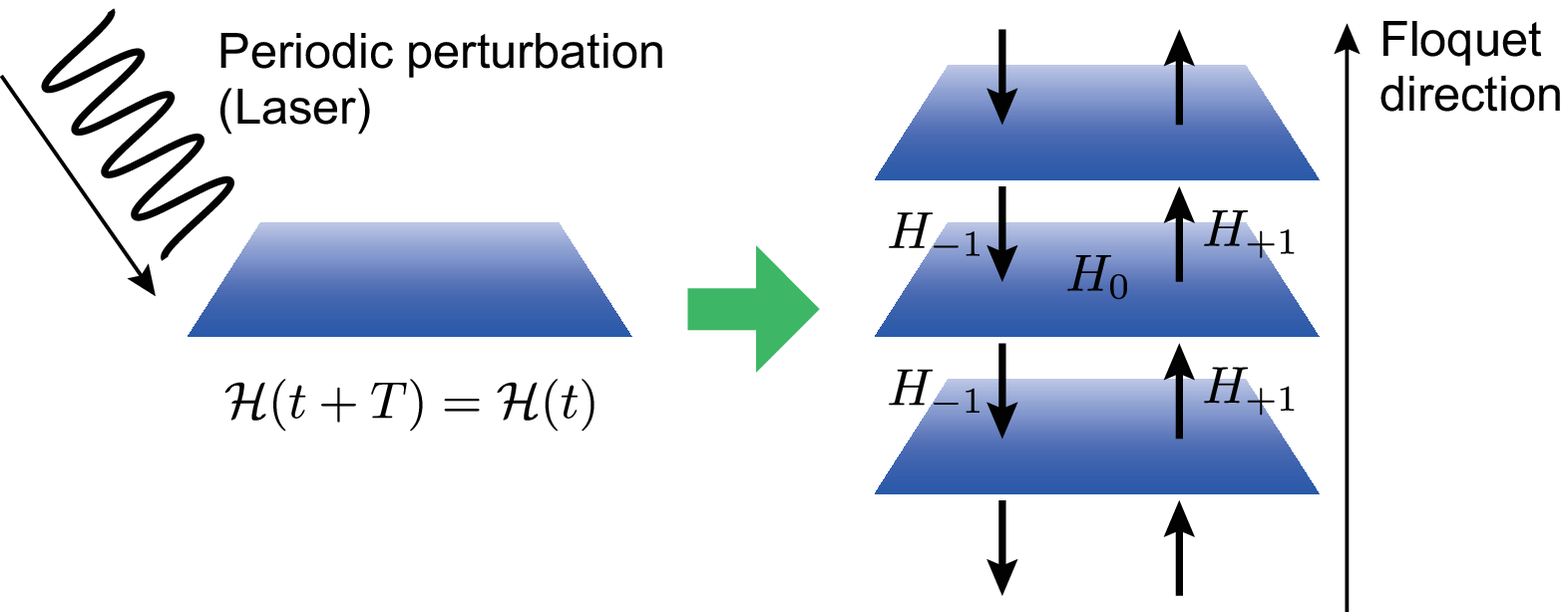}
\caption{
A system with time-periodic perturbation (left) is mapped 
to the Floquet eigenvalue problem (right). 
The latter corresponds to a system consisting of 
subspaces with different photon number (Floquet direction) coupled with each other 
through the off-diagonal components $H_{m}$ ($m\ne 0$) of the Floquet Hamiltonian.  }
\label{fig:Floquet}
\end{figure}

The infinite-dimensional eigenvalue problem Eq.~(\ref{eq:FloquetEigenMatrix}) can be reduced to a 
finite-dimensional problem using a $\Omega^{-1}$ expansion. 
If we focus on the zero photon subspace $|\Phi^{0}\rangle$, 
the neighboring subspaces $|\Phi^{\pm 1}\rangle$ 
become energetically far in the large $\Omega$ limit. 
The lowest order correction due to the off-diagonal terms $H_{\pm 1}$ 
is to simply induce virtual photon absorption-emission or emission-absorption processes
which result in an effective Hamiltonian~\cite{Kitagawa11S}
\begin{equation}
{\cal H}_{\rm eff}=H_{0}-[H_{+1},H_{-1}]/\Omega+\mathcal{O}(1/\Omega^2).
\label{eq:Effctive}
\end{equation}
This corresponds to Eq.~\eqref{eq:FloquetHeff} in the main text. 
Beyond the first order correction, either the Floquet-Magnus expansion 
or the Brillouin-Wigner perturbation can be applied. 
The comparison of these methods are given in Ref.~\cite{Mikami15}. 

As for the model considered in the main text, 
the virtual photon processes can be visualized as in 
Fig.~\ref{fig:Photon_Multiferro}. 
Let us apply this $\Omega^{-1}$ expansion scheme 
for our laser-driven system. 
In the two-spin system in Fig.~\ref{fig:Setup}(a) of the main text, 
time-dependent laser-driven terms are expressed as 
\begin{align}
{\cal H}_{\rm E}(t) =& - \boldsymbol{E}(t)\cdot \boldsymbol{P}
= -\alpha [\sin\theta\cos(\Omega t+\delta){\cal V}_{12}^z
+\cos\theta\sin(\Omega t){\cal V}_{12}^z],\\
{\cal H}_{\rm B}(t) =& -g \mu_{\rm B}\boldsymbol{B}(t)\cdot 
(\boldsymbol{S}_1+\boldsymbol{S}_2)
= \beta [\sin(\Omega t)(S_1^x+S_2^x)+\cos(\Omega t+\delta)(S_1^y+S_2^y)],
\end{align}
where $\alpha=g_{\rm me}E_{0}$ and $\beta=g\mu_{\rm B}E_{0}c^{-1}$. 
Using this expression, we can easily compute the commutator 
$[H_{+1},H_{-1}]$ and then we obtain Eq.~\eqref{eq:EffHamil} in the main text, 
\begin{equation}
{\cal H}_{\rm syn}=\frac{\alpha\beta}{2\Omega}\cos\delta
   (\boldsymbol{e}_{12}\cdot\boldsymbol{\cal V}_{12})
   +\frac{\beta^{2}}{2\Omega}\cos\delta(S_{1}^{z}+S_{2}^{z}). 
   \label{eq:EffHamilS}
\end{equation}

\begin{figure}
\centering
\includegraphics[width=0.5\textwidth]{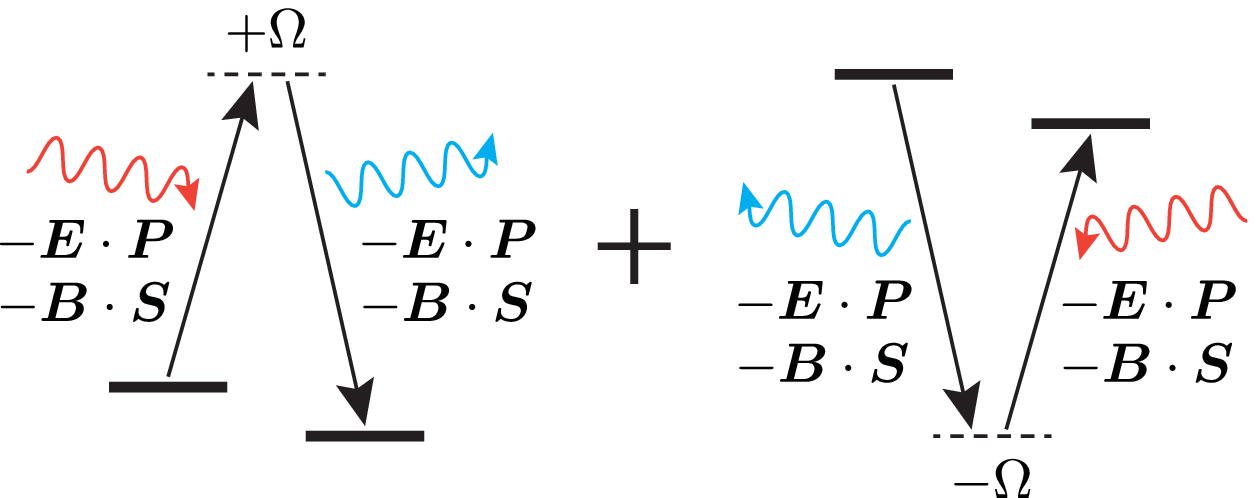}
\caption{Terms in the effective Hamiltonian (\ref{eq:EffHamilS}) 
is considered as photon absorption-emission and 
emission-absorption processes. }
\label{fig:Photon_Multiferro}
\end{figure}

\subsection{S2. Resonant-like phenomena in laser-driven multiferroic spin models}
In the main text, we showed that a resonant phenomenon 
occurs in laser-driven multiferroic spin-$\frac{1}2$ chain in 
a static Zeeman magnetic field $H$ 
with resonant frequencies around $\Omega=H$ [Fig.~\ref{fig:Field}(b)]. 
In this section, in order to show that this resonance 
is not a special feature of a particular model,
we numerically investigate laser-induced vector chirality in two other models: 
a spin-1 Heisenberg chain and a two-leg spin-$\frac{1}2$ ladder in 
circularly polarized laser with $\delta=0$. 
We assume an antisymmetric magnetostriction type
ME coupling in these two models as in the model in the main text. 
The Hamiltonian of the spin-1 chain is given by the same form as Eq.~\eqref{eq:XXZ} 
with replacing spin-$\frac{1}{2}$ operators with spin-1 ones, while that 
of the spin-$\frac{1}2$ ladder is 
\begin{align}
{\cal H}_{\rm lad}=&J\sum_{n=1,2}\sum_{j}
(\boldsymbol{S}_{n,j}\cdot\boldsymbol{S}_{n,j+1}
-\Delta S_{n,j}^{x}S_{n,j+1}^{x})
+J_\perp\sum_{j}(\boldsymbol{S}_{1,j}\cdot\boldsymbol{S}_{2,j}
-\Delta S_{1,j}^{x}S_{2,j+1}^{x})\nonumber\\
&-H\sum_{n=1,2}\sum_{j}S_{n,j}^{x}\nonumber\\
&-g_{\rm me}\boldsymbol{E}(t)\cdot\sum_{n=1,2}\sum_{j}
\boldsymbol{e}_{j,j+1}\times\boldsymbol{\cal V}_{n,j,j+1}
-g\mu_{\rm B}\boldsymbol{B}(t)\cdot\sum_{n=1,2}\sum_{j}\boldsymbol{S}_{n,j},
\end{align}
where $n$ and $j$ denote the leg and rung indices, respectively. 
The first and second terms are respectively leg ($J$) and 
rung ($J_{\perp}$) exchange couplings, 
and third is the Zeeman interaction of static external field $H$. 
The legs of the ladder are situated along the $x$ direction. 
The final line ${\cal H}_{\rm laser}(t)$ is the laser-driven time-dependent interaction, 
and $\boldsymbol{\cal V}_{n,j,j+1}=\boldsymbol{S}_{n,j}\times
\boldsymbol{S}_{n,j+1}$ is the $n$-th chain vector chirality. 
We assume that the ME coupling exists only on the leg bonds (not rung). 
The explicit form of the laser-driven term is the following: 
\begin{align}
{\cal H}_{\rm laser}(t)
=&
-g_{\rm me}E_0\sin(\Omega t)\sum_{n=1,2}\sum_{j}{\cal V}^z_{n,j,j+1}
+g\mu_{\rm B}E_0 c^{-1}\Big[
\sin(\Omega t)\sum_{n=1,2}\sum_{j}{S}^x_{n,j}
+\cos(\Omega t)\sum_{n=1,2}\sum_{j}{S}^y_{n,j}\Big]\nonumber\\
=& -\alpha \sin(\Omega t)\sum_{n=1,2}\sum_{j}{\cal V}^z_{n,j,j+1}
+\beta\Big[
\sin(\Omega t)\sum_{n=1,2}\sum_{j}{S}^x_{n,j}
+\cos(\Omega t)\sum_{n=1,2}\sum_{j}{S}^y_{n,j}\Big].
\end{align}
Using these Hamiltonians, we numerically solve their 
time-dependent Schr\"odinger equations and compute 
laser-induced vector chirality. 
The $\Omega$ dependence of time-averaged chirality is 
summarized in Fig.~\ref{fig:Chirality}, where $\Delta=0$. 
It shows that resonant peak 
structure appear just below and above $\Omega\sim H$ 
in both the spin-1 chain and spin-$\frac{1}2$ ladder 
as well as the spin-$\frac{1}2$ chain in the main text. 
These results indicate that the resonant behavior generally 
takes place in a wide class of 1D Heisenberg-type spin models with 
antisymmetric ME coupling under static magnetic field.

\begin{figure}
\centering
\includegraphics[width=0.7\textwidth]{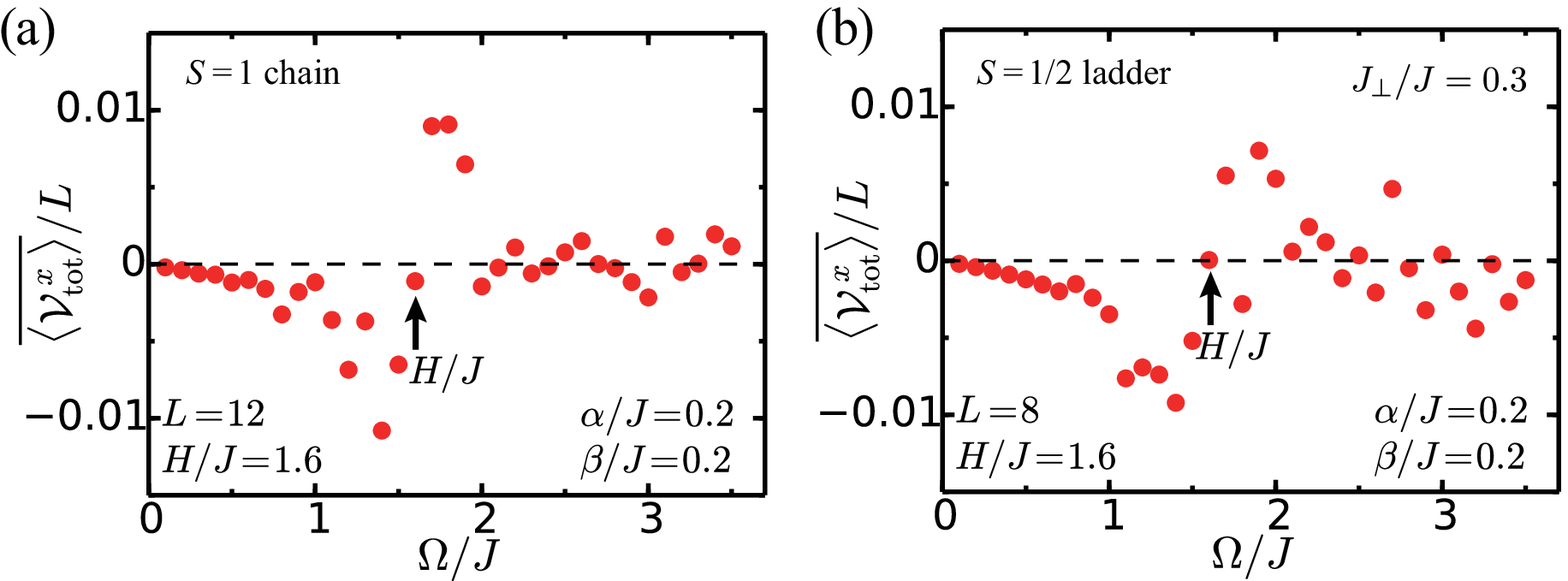}
\caption{Time-averaged vector chirality in laser-driven 
multiferroic spin-1 chain (a) and spin-$\frac{1}2$ ladder (b). 
In both the systems, resonant peaks appear just below and above $\Omega=H$. 
We choose the numerically obtained ground states of the two models 
as the initial states at $t=0$. We set $\Delta=0$ (no XXZ anisotropy), 
$H/J=1.6$, and $J_{\perp}/J=0.3$. The system size $L$ means 
the total number of sites along chain and leg directions. 
}
\label{fig:Chirality}
\end{figure}

\subsection{S3. All optical detection scheme for the laser-driven 
DM interaction and chirality}
In the main text, we explained a detection scheme of the laser-driven 
DM interaction and spin vector chirality through generating spin current. 
Here, we explain detection schemes which can be performed solely 
by an optical setup. For systems with a ferromagnetic order, 
time-resolved magnetooptical Kerr effect (MOKE) can be used to study the 
dynamical change of magnetization. Figures~\ref{fig:Detection}(a) and (b) 
show setups for measuring laser-driven 
DM interaction through Faraday and Kerr effects, respectively. 
As for the pump, THz laser is preferred since the energy scale of excitations 
in multiferroic magnets typically corresponds to this frequency region.
When we apply a suitable THz circularly 
(or elliptically) polarized laser to multiferroic ferromagnets, 
its magnetic order is changed from a uniform ferromagnet to a spiral order 
due to the dynamical DM interaction. This leads to a decrease of 
the uniform magnetization. 
Such demagnetization can be detected through the change in the Faraday 
or Kerr rotation angle before and after the application of pump laser.

A more direct detection scheme is through 
directional dichroism [Fig.~\ref{fig:Detection}(c)]. 
This is a magnetooptical phenomenon 
where the light transmission becomes directionally dependent. 
As we mentioned above, a noncollinear order is generally expected to 
emerge when circularly (or elliptically) polarized laser is applied to 
multiferroic, ferro or antiferro-magnets 
and such a spiral state can lead to 
directional dichroism~\cite{Miyahara12,Takahashi14}. 
Therefore, observing directional dichroism via probe laser 
would be a smoking gun experiment for laser-driven DM interaction.

Finally, we discuss how to distinguish phenomena induced by synthetic 
DM interaction from other laser-driven effects. 
Varying the laser frequency $\Omega$ would provide a way of the distinction. 
As we mentioned in the main text, since the ME coupling $g_{\rm me}$ 
is strong within the GHz to THz regime, the synthetic DM interaction 
cannot emerges in other frequency regimes, e.g., optical regime, 
while heating effect would not be so sensitive 
to the laser frequency $\Omega$. 
A more precise way to distinguish the mechanisms is to change the incident 
direction of the pump laser. Equation~\eqref{eq:KNB} in the main text 
\begin{equation}
\boldsymbol{P}=g_{\rm me}\boldsymbol{e}_{12}\times
   (\boldsymbol{S}_{1}\times\boldsymbol{S}_{2}),
\end{equation}
shows that the ME interaction between polarization and pump laser 
strongly depends on the geometric relation between the crystal axis 
$\parallel {\boldsymbol e}_{12}$ and the laser direction. 
The strength of laser-driven DM interaction is thereby changed when 
varying the incident direction of the pump laser and this will be generally different from other laser-driven phenomena including heating effect.

\begin{figure}
\centering
\includegraphics[width=0.7\textwidth]{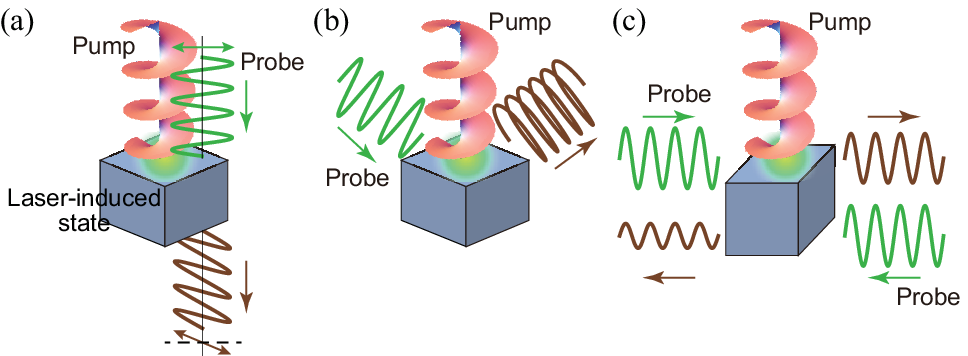}
\caption{Schematic images of (a) Faraday effect, (b) Kerr effect, 
and (c) directional dichroism in the presence of pump laser being 
circularly polarized. Signatures of laser-driven DM interaction 
and spin chirality are expected to be detected by measuring these 
magnetooptical effects.
}
\label{fig:Detection}
\end{figure}

\subsection{S4. Spatially modulated laser and spin currents}
In the final part of the main text, we discussed a way of generating spin 
current by applying spatially modulated laser which can be 
induced in near field of chiral plasmonic structures.
To confirm that a spatial modulation of the laser results in an 
inhomogeneous chirality and thus to spin current generation, 
we perform numerical simulations 
for multiferroic spin-$\frac{1}2$ chains with a spatially modulated laser. 
The effect of spatial modulation is incorporated by making 
the field strength parameters $\alpha$ and $\beta$ site-dependent as 
\begin{equation}
\alpha_{j,j+1}=\alpha(\sin^{2}(\pi j/L)+\sin^{2}(\pi(j+1)/L))/2, \quad
\beta_{j}=\beta\sin^{2}(\pi j/L),
\label{eq:modulation}
\end{equation}
respectively. 
The numerical results are shown in Fig.~\ref{fig:SpaceModulation}. 
The driven vector chiralities strongly depend on their site positions, 
which clearly indicates an emergence of a finite spin current 
$\langle dS_{j}^{x}/dt\rangle$.

\begin{figure}
\includegraphics[width=0.7\textwidth]{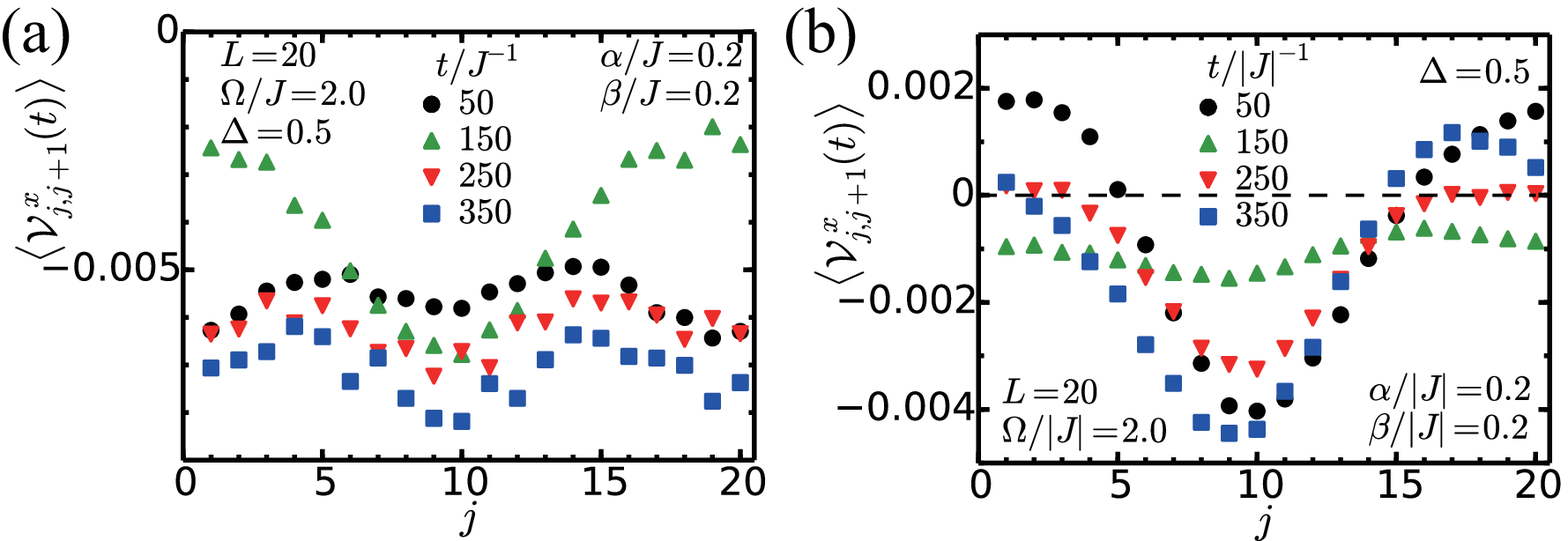}
\caption{Time evolution and site dependence of 
$\langle{\cal V}_{j,j+1}^{x}(t)\rangle$ in (a) AF and 
(b) ferromagnetic XXZ models under a spatially modulated laser with 
$\alpha_{j,j+1}$ and $\beta_{j}$.}
\label{fig:SpaceModulation}
\end{figure}

\end{widetext}

\end{document}